# CONSEQUENCES OF KALUZA-KLEIN COVARIANCE

Paul S. Wesson

Department of Physics and Astronomy, University of Waterloo, Waterloo, Ontario N2L 3G1, Canada

Space-Time-Matter Consortium, http://astro.uwaterloo.ca/~wesson

Addresses: Mail to Waterloo above; email: psw.papers@yahoo.ca

Abstract

The group of coordinate transformations for 5D noncompact Kaluza-Klein theory is broader than the 4D group for Einstein's general relativity. Therefore, a 4D quantity can take on different forms depending on the choice for the 5D coordinates. We illustrate this by deriving the physical consequences for several forms of the canonical metric, where the fifth coordinate is altered by a translation, an inversion and a change from spacelike to timelike. These cause, respectively, the 4D cosmological 'constant' to become dependent on the fifth coordinate, the rest mass of a test particle to become measured by its Compton wavelength, and the dynamics to become wave-mechanical with a small mass quantum. These consequences of 5D covariance – whether viewed as positive or negative – help to determine the viability of current attempts to unify gravity with the interactions of particles.

1. Introduction

Covariance, or the ability to change coordinates while not affecting the validity of the equations, is an essential property of any modern field theory. It is one of the founding principles for Einstein's theory of gravitation, general relativity. However, that theory is four-dimensional, whereas many theories which seek to unify gravitation with the interactions of particles use higher-dimensional spaces. The question then arises of what happens to 4D physical quantities when the higher-dimensional coordinates on which they depend are changed. We examine this problem briefly here, for modern five-dimensional theories of the Kaluza-Klein type. (We recognize, of course, that a complete unified field theory which incorporates the internal symmetry groups of particle physics will probably require more than 5 dimensions.) In 5D, the group of coordinate transformations $x^A \to \bar{x}^A(x^B)(A = 0-4)$ is broader than the 4D group $x^\alpha \to \bar{x}^\alpha(x^\beta)$ $(\alpha = 0-3)$. It is therefore possible for a 4D quantity $Q(x^\alpha, x^4)$ to change its form under a coordinate transformation which includes the extra dimension. The consequences of this provide a way to assess the viability of 5D theories.

The original unification of gravitation and electromagnetism by using 5 dimensions was in 1921 by Kaluza [1], but he neglected the effect of the fifth or scalar potential, and set all derivatives of 4-dimensional quantities with respect to the fifth coordinate to zero (the 'cylinder' condition). An attempt to bring in quantum effects was made in 1926 by Klein [2], but his identification of the electron charge with the momentum of a particle in the fifth dimension meant that the latter was rolled up into an unobservably small circle ('compactification'). A different approach was made by Dirac in 1935 [3], who game a neat if unfamiliar classification of 4D particle properties such as energy and momentum in terms of an embedding in a 5D spherical space. Modern versions of Kaluza-Klein theory allow the fifth coordinate to play an important physical role, and the fifth dimension is not compact and can indeed be large. Space-time-matter theory dates from 1992, and is motivated by the old idea of Einstein to give a geometrical description of matter [4, 5]. It does this by embedding the 4D Einstein equations with sources in the apparently-empty 5D Ricci-flat equations, so there is an effective or induced energy-momentum tensor in which properties of matter like the density and pressure are given in terms of the fifth potential and derivatives of the 4D spacetime po-tentials with respect to the fifth coordinate. Membrane theory dates from 1998, and is motivated by the wish to explain the strength of particle interactions compared to gravity, or alternatively the smallness of particle masses compared to the Planck value [6, 7]. It does this by embedding 4D spacetime as a singular hypersurface in a 5D manifold, so particle interactions are confined to a sheet while gravity is diluted by propagating into the bulk of the fifth dimension. Both space-time-matter theory (sometimes called induced-matter theory) and membrane theory are in agreement with observations.

Space-time-matter (STM) theory is fully covariant in 5D. It is based on manifestly coordinate-invariant field equations which involve the 5D Ricci tensor, namely $R_{AB} = 0 \, (A, B = 0-4)$. These include as a subset the 4D equations of general relativity which involve the Einstein tensor and the energy-momentum tensor, namely $G_{\alpha\beta} = 8\pi T_{\alpha\beta}$ $(\alpha, \beta = 0-3)$. Here, we are labelling the time, space and extra coordinates via $x^0 = t$, $x^{123} = xyz$ and $x^4 = l$, where the last appellation is to avoid confusion with the Cartesian measure, and the implication that the fifth coordinate is measured from some singular hypersurface. Membrane (M) theory employs the latter to define spacetime, which is usually located at $x^4 = 0$. (Particles are confined to this hypersurface in M theory because it is singular, whereas in STM theory particles may drift with respect to a given hypersurface, but at a slow rate given by the cosmological constant; see below.) The presence of a special surface in the manifold means that M theory has a restricted form of 5D covariance. In STM theory, the 5D metric is commonly taken in the canonical form, where spacetime is multiplied by a quadratic factor in $x^4$ and there is an additional flat part. In M theory, the 5D metric is commonly taken in the warp form, where spacetime is multiplied by an exponential factor in $x^4$ and there is an additional flat part. In both metrics, the $x^4$-dependent prefactor on spacetime can be changed, in accordance with covariance. This is usually associated with some application of the metric to a physical system. The general question we therefore wish to pose is: What happens to 4D-measurable quantities when the 5D coordinates (especially $x^4$) are changed? This question is of particular relevance to STM theory. Mathematically, its approach is justified by a local embedding theorem from 1926 due to Campbell [8]. Physically, this means that quantities such as the density $\rho$ and pressure $p$ of a fluid, or the rest mass $m$ of a particle, are specified by the equations as functions of $x^\alpha$ and $x^4 = l$. For example, there is a large body of results which indicate that when the 5D metric of STM-theory has the canonical form, it gives back conventional mechanics with the identification $l = m$. (Here we are absorbing the gravitational constant $G$, the speed of light $c$ and Planck's constant $h$ by a choice of units which renders them unity, though we will restore these physical parameters when relevant.) It is obviously important to ask what happens to the 4D particle rest mass when the form of the canonical metric is changed by a coordinate transformation that involves $x^4 = l$. Some results on this and similar issues are available in the literature [9, 10]. We wish in what follows to give a more systematic account of the nature of mass in 5D field theory.

We will use the established nature of the canonical coordinate system, but inquire what happens when its extra coordinate $x^4 = l$ undergoes a translation $(l \to l + l_0)$, an inversion $(l \to 1/l)$, and a change that converts the extra dimension from spacelike to timelike $(l \to il)$. We will confirm that the first change converts the cosmological constant from a true constant to an *l*-dependent function, which can help in resolving the cosmological-'constant' problem [11, 12]. The second change will be seen to correspond to a kind of units transformation, in which a mass goes from being measured by its Schwarzschild radius $(Gm/c^2)$ to being measured by its Compton wavelength $(h/mc)$. In accordance with this, the third change will be found to correspond to a shift in the nature of the dynamics from classical to wave-like, with an associated quantum of mass which is too small to have been detected to date but can in principle be measured. These three results provide a means of assessing the viability of STM theory and similar higher-dimensional unified field theories.

2. Gauge-Dependent Physics

We wish to briefly review properties of the canonical metric, prior to deriving some specific consequences of it in the next section. This metric has been much studied, and we therefore employ it to illustrate how 4D physics can depend on a change of 5D coordinates or gauge.

That there can be gauge-dependent effects is clear from the aforementioned fact that the group of coordinate transformations in 5D is broader than the group in 4D. Even in special relativity, expressions for quantities such as the energy and momentum of a test particle depend on whether the reference frame is moving or not. In a covariant 5D ver-sion of general relativity, any quantity $Q(x^\alpha, l)$ will alter its form depending on the choice of $x^4 = l$. That said, there are of course invariants in 5D relativity as there are in 4D. One such is the interval, which in 5D is defined by $dS^2 = g_{AB} dx^A dx^B$ ($A, B = 0-4$), and contains the usual 4D one $ds^2 = g_{\alpha\beta} dx^\alpha dx^\beta$ ($\alpha, \beta = 0-3$). Due to the form of the 5D interval (see below), it is possible that massive particles travelling on timelike paths with $ds^2 > 0$ in 4D can be travelling on null paths with $dS^2 = 0$ in 5D. This is the case in both STM theory [13] and M theory [14]. The null condition of $dS^2 = 0$ has been extensively used in studies of 5D dynamics, because useful relations can then be obtained directly from the metric, without the need to solve the 5D geodesic equation (though the latter has also been well studied, as in refs. 9-11). We will therefore in what follows adopt the view that massive particles in 4D travel on null paths in 5D.

The line element for the canonical metric is commonly written

$$dS^2 = (l/L)^2 g_{\alpha\beta}(x^\gamma, l) dx^\alpha dx^\beta - dl^2 \qquad (1.1)$$

$$= (l/L)^2 ds^2 - dl^2 \qquad . \qquad (1.2)$$

This is algebraically general, because it uses the 5 available degrees of coordinate freedom to remove the terms $g_{\alpha 4}$ which correspond to the electromagnetic potentials ($A_\alpha \equiv g_{\alpha 4}/g_{44}$) and to flatten the term $g_{44}$ which describes the scalar field ($g_{44} = -\Phi^2$). Both of these steps can be reversed by appropriate gauge changes, but these go beyond the scope of the present study. The factorization by $(l/L)^2$ in (1) involves a constant length $L$, and from a mathematical viewpoint is done to establish a coordinate frame analogous to the synchronous one of standard cosmology. That is, in (1) all observers agree on the value of $x^4 = l$. However, the factorization in (1) also has a physical meaning, in that its first part gives back the standard element of action $mds$ if $l = m$ in the appropriate limit. That is, in (1) the first part involves a geometrical description of the rest mass $m$ of a test particle, so we effectively have a momentum manifold rather than a coordinate manifold. To verify this, it can be shown that the constant of the motion associated with the time axis of (1) in the appropriate limit where the 3-velocity is $v$ is just the standard $m(1-v^2)^{-1/2}$ if $l = m$ [9, 15]. In regard to the geometrical description of mass in the canonical metric (1), it should be noted that in general it is not constant in 5D relativity. This should not be surprising, because there are cases even in 4D physics where the mass varies. One such is a rocket, which loses mass as it burns fuel, and thereby gains velocity. In other words, the essential thing in the local limit is the conservation of momentum. To investigate this, we recall that for a massive particle the 4-velocities $u^\alpha \equiv dx^\alpha / ds$ in conventional dynamics are normalized via the condition $g_{\alpha\beta}(x^\gamma) u^\alpha u^\beta = 1$. The corresponding condition in 5D is $g_{\alpha\beta}(x^\gamma, l) u^\alpha u^\beta = 1$, which can be varied with respect to the extra coordinate $l$ and the 4D proper time $s$ and usually produces a term in $dl/ds$. This represents a motion between the 4D and 5D frames, which in general will produce an inertial force as measured in 4D. This force has been isolated in both STM and M theory [15, 16]. It is really an acceleration, which acts parallel to the 4-velocity, and is given by $P^\mu = (-1/2) (\partial g_{\alpha\beta}/\partial l) u^\alpha u^\beta (dl/ds) u^\mu$. The equation of motion in the local limit then reads $du^\mu / ds = P^\mu$. This for the canonical metric (1) gives $d(lu^\mu)/ds = 0$, which expresses conservation of momentum along the spacetime path when $l = m$. More complicated applications of this so-called fifth force have been discussed elsewhere [17]. Here, we just need to note that the geometrization of mass in the metric (1) via $x^4 = l = m$ is consistent with the law of conservation of momentum.

The canonical metric (1) is convenient because it allows us to identify the extra coordinate $x^4 = l$ in terms of the rest mass $m$ of a test particle via a comparison with established dynamics. In this manner, the common physical labels for a test body ($t$, $xyz$ and $m$) are treated on the same footing in a space-time-mass manifold. However, the underlying theory is covariant, so these labels can be changed at will. And in general, an arbitrary coordinate transformation or gauge change for (1) will result in a metric where the correspondence between $l$ and $m$ becomes obscure. Conversely, a solution of the 5D field equations $R_{AB} = 0$ in arbitrary coordinates may have to be subjected to a gauge change towards (1) if its mass-related physics is to be interpreted in conventional terms. The same situation occurs in 4D for general relativity. For example, had the Schwarzschild solution been discovered in Kruskal-Szekeres coordinates, a complicated change of gauge would have been necessary to bring the metric into a form applicable to the solar system [18]. These comments apply not only to the dynamical properties which follow from the metric, but also to the properties of matter for a fluid which follow from the field equations. The 5D group of coordinate transformations $x^A(x^B) \to \bar{x}^A(x^B)$ preserves $R_{AB} = 0$. The 4D group $x^\alpha(x^\beta) \to \bar{x}^\alpha(x^\beta)$ preserves $G_{\alpha\beta} = 8\pi T_{\alpha\beta}$. It is the difference between these which lies behind the fact that Birkhoff's theorem does not hold in 5D, and that there is more than one 5D 'vacuum' [19]. In general relativity, the vacuum is usually regarded as the state where there is no 'ordinary' matter, and is specified by a unique parameter, namely the cosmological constant $\Lambda$. This measures the energy density and pressure of the Einstein vacuum if it is regarded as a fluid and included in the energy-momentum tensor. Then its equation of state is $p_v = -\rho_V = -\Lambda/8\pi$ [17-19]. In noncompact Kaluza-Klein theory with the canonical metric (1), the absence of ordinary matter is known to correspond to $\partial g_{\alpha\beta}/\partial l = 0$ [11, 17]. This condition leads via the field equations to a physical identification of the constant length $L$ in (1) in terms of the conventional cosmological constant. Namely, $\Lambda = 3/L^2$. We will comment further on this below, but here we note that in (1) a spacelike extra coordinate implies $\Lambda > 0$ while a timelike one implies $\Lambda < 0$. Also, the condition $\partial g_{\alpha\beta}/\partial l = 0$ in (1) leads via the equations of motion to a recovery of the conventional Equivalence Principle [19], which can be regarded as a 5D geometric symmetry.

The Equivalence Principle in STM theory is actually related to the hierarchy problem (or the small values of the masses of real particles as compared to the theoretical Planck value), which was noted above in connection with M theory. For the (weak) Equivalence Principle states that the gravitational mass of an object is *proportional* to its inertial mass. In STM theory, these masses are geometrized via the Schwarschild radius $l_g \equiv Gm_g/c^2$ and the Compton wavelength $l_i \equiv h/m_ic$. So the noted proportionality reads $m_g \sim m_i$ or $(c^2l_g/G) \sim (h/cl_i)$ or $l_gl_i \sim Gh/c^3$. Introducing a constant length $L$, this proportionality can be written as the equation $l_gl_i = L^2$, where however we are *not* obliged to identify L with the Planck length. To do so precipitates the hierarchy problem, with the particle mass being necessarily the Planck one $(hc/G)^{1/2} \approx 10^{-5} g$, which is unacceptably large. In other words, STM avoids the hierarchy problem by recognizing that $l \to L^2/l$ is a gauge change between gravitational and inertial measures of mass, with distinct choices of coordinate. We will investigate this in detail below. (The physical dimensions of our equations will always balance, provided we are clear about whether we use the so-called Einstein or Planck gauges and do not mix them; see refs. 9, 11.) A possible criticism of this approach to mass is that it is restricted to spinless, uncharged particles. For such particles, there is a generic class of solutions of the field equations, with different numerical parameters corresponding to different numerical values of the mass [5, 17]. No significant exact solutions of the field equations are yet known which incorporate spin. However, this may be viewed as a merely technical problem, because Campbell's theorem ensures that a 5D analog of the 4D Kerr solution must exist. The situation as regards electric charge is more satisfactory, because exact solutions are known, and their dynamics have been investigated (see ref. 17, pp. 169-180). With respect to dynamics, these have been studied using the geodesic equation, the Lagrangian and the Hamilton-Jacobi approach [5, 9, 10, 11, 17]. The last has been examined in particular by Ponce de Leon [10], and is appropriate if there is a 5D concept of particle mass. However, in the original STM approach, there is no 5D mass as such, and the 4D parameter is a result of the geometry, in a manner analogous to how the matter density is induced from the higher-dimensional geometry. In the case where the 5D manifold is empty, and mass and density are 4D quantities related to the geometrical embedding as constrained by Campbell's theorem, the appropriate dynamics should logically be based on the 5D interval being null. This is the simplest approach, and since it works for both STM and M theory [13, 14] is the one adopted here.

In view of the above comments, coordinate changes to the canonical metric (1) should be approached with caution, if we are to retain physical insight. For this reason, in what follows we study 3 relatively minor gauge changes, all restricted to the extra coordinate.

3. Three Gauge Changes and Their Consequences

In this section, we will carry out three simple changes to the fifth coordinate $l$ and inquire how these alter our view of physics in 4D spacetime.

The change $l \to (l - l_0)$ appears almost trivial, and is so as regards the mass $m$, which is merely shifted along the $x^4$ axis. This leaves the last term in the canonical metric (1) unaltered. However, it alters the first term, which we saw previously is connected with the 4D cosmological constant via $\Lambda = 3/L^2$, where the constant length $L$ fixes the 4D curvature. As measured by the 4D Ricci scalar for a vacuum spacetime, $|R| = 4|\Lambda|$, where both signs for $\Lambda$ are admissible depending on the signature (see above). In a covariant 5D theory like STM, a 4D quantity such as $R$ can be measured in two ways: intrinsically in the hypersurface we call spacetime, and extrinsically in the orthogonal fifth dimension. An analogy is with the surface of the Earth, whose curvature can be determined intrinsically by mapping triangular figures over its surface, or extrinsically in terms of the radius measured from the centre. In 5D, both measures involve an analysis of the field equations, the details of which are already in the literature [5, 11, 17]. For metric (1), the value $\Lambda = 3/L^2$ quoted above is the intrinsic measure. It is arrived at by evaluating the 4D Einstein tensor in the case where the 4D part of the 5D metric tensor is written as $\gamma_{\alpha\beta}(x^\mu, l) = (l/L)^2 g_{\alpha\beta}(x^\mu \text{ only})$ and $g_{44} = -1$, so specifying the vacuum. We could, alternatively, calculate $\Lambda$ extrinsically, using the expression for the 4D Ricci scalar evaluated in terms of $x^4 = l$ when $|g_{44}| = 1$ as before. The relevant expression is

$$|\Lambda| = \frac{|R|}{4} = \frac{1}{16} \left| \frac{\partial \gamma^{\alpha\beta}}{\partial l} \frac{\partial \gamma_{\alpha\beta}}{\partial l} + \left( \gamma^{\alpha\beta} \frac{\partial \gamma_{\alpha\beta}}{\partial l} \right)^2 \right| \quad , \tag{2}$$

where there is summation. For $\gamma_{\alpha\beta} = (l/L)^2 g_{\alpha\beta}(x^\mu \text{ only})$ this gives $\Lambda = 3/l^2$ for space-like $l$. The apparent discrepancy between this value and the one $\Lambda = 3/L^2$ noted previously is merely the prefactor $(l/L)^2$ in (1), and reflects the difference between the two measures. Both appear in the literature, and the difference depends on whether $\Lambda$ is measured by methods confined to spacetime or ones which can go outside it. This is itself an example of how 5D covariance can impact 4D physics, though only a minor one. A more significant result follows when we implement the shift $l \to (l - l_0)$ and evaluate $\Lambda$ intrinsically by recalculating the relevant quantities in the spacetime hypersurface. The working for this can be done in different ways, which while tedious lead to the same result [11]. We quote the answer for the 'pure' canonical metric (meaning the form with $\partial g_{\alpha\beta}/\partial l = 0$ which preserves the Equivalence Principle):

$$dS^2 = \left(\frac{l-l_0}{L}\right)^2 g_{\alpha\beta}(x^\gamma) dx^\alpha dx^\beta - dl^2 \qquad (3.1)$$

$$\Lambda = \frac{3}{L^2}\left(\frac{l}{l-l_0}\right)^2 \quad . \qquad (3.2)$$

This expression has interesting consequences for physics. Primarily, the cosmological 'constant' of Einstein theory is seen to depend on the fifth coordinate, and only has its standard value $3/L^2$ in the limit $l \to \infty$. In general, it is a function of $x^4 = l$, and has a magnitude dependent on the local value of that coordinate as measured at some place in spacetime. (In general, we expect $l = l(s)$ unless there is confinement by nongravitional forces to a given hypersurface.) The cosmological 'constant' can even diverge for $l \to l_0$ and provide thereby a model for the big bang, with a decaying vacuum energy that is compatible with recent astrophysical data [20]. Alternatively, (3) can be applied to local particle physics in the interpretation where $l$ measures rest mass $m$ (see above), to infer $\Lambda$ has a size much larger on small scales than the tiny (mean) value measured on cosmological scales. Indeed, if $E = (l - l_0)$ is taken as a measure of energy, then (3) implies that $\Lambda E^2 \simeq$ constant in an interaction, a relation that should be testable using the Large Hadron Collider. In summary, the gauge change $l \to (l - l_0)$ is seen to provide a new view of the cosmological 'constant'. It has been known for a while that the values of $\Lambda$ as measured on large and small scales are discordant by a factor in the range $10^{60}$ - $10^{120}$ [12]; and our preliminary investigations show that in principle this problem can be resolved in terms of 5D gauge choices.

In the preceding, we identified the extra coordinate $x^4 = l$ for the canonical metric (1) in terms of the rest mass *m*, using dynamical arguments. The algebraic identification $l = m$ presupposes that there are fundamental constants available which makes such a choice acceptable also in a physical sense. This is of course the case; and with the gravitational constant and the speed of light, we have implicitly been measuring the mass of a test particle in terms of its Schwarzschild radius $Gm/c^2$. However, this is not the only way to geometrize mass which is suggested by the constants of nature. An alterative involves the quantum of action and is in terms of the Compton wavelength $h/mc$. From the viewpoint of physics, these two ways to parametize the mass of a test particle as a length are actually unique. They are typically employed in the classical and quantum domains. The difference is sometimes discussed as one between gravitational and atomic units, or as the Einstein versus the Planck gauge. In the present context, the difference is simply another choice of 5D gauge.

The change $l \to L^2/l$ causes the (pure) canonical metric to take on the form

$$dS^2 = (L/l)^2 ds^2 - (L/l)^4 dl^2 \quad . \tag{4}$$

This, for the 5D null path $dS^2 = 0$ as discussed above, implies

$$dl/ds = \pm(l/L) \quad . \tag{5}$$

This result also follows from the original canonical metric (1) when the 5D path is null. However, other consequences of (4) are not the same as for (1). The identification for the mass, either by the action implied by (4) or other means, is now via $l = 1/m$ (with con-stants absorbed). Then (4) with $dS^2 = 0$ gives a relation between physical and geometrical quantities which reads

$$mds = \pm d(L/l) \quad . \tag{6}$$

The implication is clearly that the conventional action $mds$ is observed to be quantized because $L/l = n$, an integer. This is reminiscent of the original Klein model [2], in which the fifth dimension is compactified to a circle, so by (5) the electron charge is proportional to $dl/ds = \pm(1/n)$ and is also quantized. However, we should be wary of drawing conclusions about the topology of the fifth dimension, because (6) as it stands says only that

$$\int mcds = nh = (L/l)h \quad . \tag{7}$$

This explains physical quantization in 4D as a consequence of the algebraic constraint $L/l = n$ in 5D, but does not explain the origin of the latter condition. [Note that in 4D theories with a scalar field and in 5D STM, the mass generally varies along the path via $m = m(s)$, so in (7) the mass has to be kept inside the integral.] One interesting implication of (7) is that there is a kind of quantum of mass, determined by $n = 1$ and the value of $L$ indicated by the cosmologically-measured $\Lambda = 3/L^2$. It is

$$m = (h/c)(\Lambda/3)^{1/2} \simeq 2 \times 10^{-65} g \quad . \tag{8}$$

This is tiny, and explains why mass appears to be unquantized in experiments. The nature of a mass quantum like (8) has been discussed elsewhere [9], so we leave the subject here, and turn to another gauge choice which is compatible with quantization.

The change $l \to il$ with $L \to iL$ causes the canonical metric to effectively change signature, from $(+----)$ to $(+---+)$. That is, a Wick rotation of the extra coordinate and a corresponding change in the associated lengthscale of the potential changes the fifth dimension from spacelike to timelike. (There is no problem with closed timelike paths, because the fifth coordinate is not really a time but is related to mass.) The 5D null path is now given by

$$dS^2 = 0 = (l/L)^2 ds^2 + dl^2 \quad . \tag{9}$$

This has the wavelike solution $l = l_* \exp(\pm is/L)$, which describes motion with wavelength $L$ and amplitude $l_*$ around $l = 0$. If a shift by $l_0$ is included as before, the motion becomes

$$l = l_0 + l_* \exp(\pm is/L) \quad . \tag{10}$$

In principle, this expression can be used to calculate the wavelike properties of a particle in the classic double-slit experiment and neutron interferometry [21]. However, this requires physical assumptions about the relationship between $l$ and $m$, and about the topology of the fifth dimension (see above). It also requires certain mathematical assumptions. (For example, the common practice of using a cosine wave to represent the real part of a complex quantity is strictly only valid in the linear case, whereas the interval which underlies the preceding analysis is quadratic.) We therefore defer a detailed investigation of (10), and here confine our discussion to three comments aimed at indicating the potential utility of the gauge (9). First, the question of confinement is automatically answered because the motion is oscillatory about a hypersurface. Second, the mass is in general wavelike, so a particle's energy and 3-momentum are also wavelike, in the manner of deBroglie. Thirdly, the mass may consist of real and imaginary parts, as in the original scheme of Dirac [3] mentioned at the beginning of this account.

4. Conclusion

The coordinates with which we describe the world can be chosen for convenience but are essentially arbitrary; so we ensure that the equations concerned retain their validity for all choices of coordinates, and establish the Covariance Principle. This and the Equivalence Principle are the foundation of 4D general relativity. However, there is nothing in the Einstein equations which limit their dimensionality; and the best route to a unification of gravity with the interactions of particle physics would appear to be via extra dimensions. Unfortunately, the original form of 5D relativity due to Kaluza and Klein, while it successfully unified gravity with electromagnetism, was hobbled by the cylinder and compactification conditions, which restricted its usefulness and led ultimately to its abandonment. (The cylinder condition, where all derivatives with respect to the extra coordinate were set to zero, is plainly non-covariant.) Currently, we have two more successful forms of 5D relativity: space-time-matter theory and membrane theory. The latter has a singular hypersurface, which somewhat limits its covariance; but the former is fully covariant. (The independence of the 4D metric tensor from the extra coordinate when the line element has the canonical form is a symmetry that implies the Weak Equivalence Principle, and may be violated at some level.) The field equations of STM theory are the 5D Ricci-flat ones $R_{AB} = 0 \ (A,B = 0-4)$. These contain the 4D Ein-stein equations with a matter source which is basically geometric in origin, a result ultimately of Campbell's embedding theorem [8]. Numerous exact solutions of the theory are known, including many which depend on the extra coordinate [17]. These solutions, plus the form of the field equations, necessarily imply that 4D physics as restricted to some hypersurface (spacetime) will appear to be dependent on the fifth coordinate, i.e. gauge-dependent.

This gauge dependence of 4D physics may be viewed either as an advantage or as a drawback. It is an advantage insofar as it widens the scope of physics, giving us explanations of quantities such as the rest mass and electric charge of a test particle. It may be viewed as a drawback insofar as these and other quantities may now be variable whereas before they were constant. In general, a particle in a 5D manifold will move away from some 4D hypersurface unless constrained by nongravitational forces (4D null geodesics are exceptions). For the canonical metric with which we have been mainly concerned, this means that when the extra coordinate is identified with rest mass, the latter varies with 4D proper time, though at a slow rate governed by the cosmological 'constant'. If this effect is not desired, it may be suppressed, either by the introduction of a singular membrane [6], or by restricting the group of coordinate transformations used to define spacetime-observable quantities [10]. However, both of these compromise 5D covariance.

We have in the above looked at what happens when we stick with covariance, and carry out three simple changes to the extra coordinate when the metric has the canonical form (1). A *shift* along the extra axis changes the cosmological 'constant' to a function (3), suggesting that each particle has its 'own' value of the parameter. This suggestion can help resolve the mismatch in the energy of the vacuum as measured locally and globally [12, 13]. For particle physics, it implies that the product of the local value of the cosmological 'constant' and the square of the energy is constant for an interaction. An *inversion* of the extra coordinate leads immediately to a recovery of the standard rule for the quantization of the action in terms of structure in the extra dimension. This is seen in equations (4)-(8), which are typical results for what in the literature is sometimes called the Planck gauge, as opposed to the classical Einstein gauge. Lastly, a change to an *imaginary* extra coordinate effectively gives a 5D metric with a timelike extra dimension (9), in which a particle is confined near to a hypersurface and oscillates around it as in equation (10). That is, a particle becomes a wave, and preliminary calculations indicate that this gauge is also compatible with quantization. In conclusion, all three gauges have major implications for our understanding of the nature of mass.

The consequences of covariance discussed here are, in a physical sense, remarkably far-reaching. They may be viewed as positive or negative. But in either case, they provide ways to test the viability of five-dimensional relativity.

Acknowledgements

Thanks for the comments on 5D go to members of the S.T.M. Consortium, and for past discussions on neutron interferometry to S. Werner. This work was partly supported by N.S.E.R.C.

References

[1] Kaluza, T. 1921 *Sitz Preuss. Akad. Wiss.* **33** 966.

[2] Klein, O. 1926 *Z. Phys.* **37** 895.

[3] Dirac, P.A.M. 1935 *Ann. Math.* **36** 657.

[4] Wesson, P.S. 1992 *Phys. Lett B* **276** 299. Wesson, P.S., Ponce de Leon, J. 1992 *J. Math. Phys.* **33** 3883. Wesson, P.S. 1992 *Astrophys. J.* **394** 19.

[5] Wesson, P.S. 2008 *Gen. Rel. Grav.* **40** 1353.

[6] Randall, L., Sundrum, R. 1998 *Mod. Phys. Lett. A* **13** 2807. Arkani-Hamed, N., Dimopoulos, S., Dvali, G. 1998 *Phys. Lett. B* **429** 263.

[7] Maartens, R. 2004 *Living Reviews Relativity* **7** 1, gr-qc/0312059.

[8] Campbell, J.E. 1926 *A Course of Differential Geometry* (Clarendon: Oxford). Seahra, S.S., Wesson, P.S. 2003 *Class. Quant. Grav.* **20** 1321. Dahia, F., Romero, C., Da Silva, L.F.P., Tavakol, R. 2007 gr-qc/0702063. Dahia, F., Romero, C., Da Silva, L.F.P., Tavakol, R. gr-qc/07111279.

[9] Wesson, P.S. 2002 *J. Math. Phys.* **43** 2423. Wesson, P.S. 2002 *Class. Quant. Phys.* **19** 2825.

[10] Ponce de Leon, J. 2001 *Mod. Phys. Lett. A* **16** 2291. Ponce de Leon, J. 2002 *Int. J. Mod. Phys. D* **11** 1355. Ponce de Leon, J. 2002 *Grav. Cosmol.* **8** 272. Ponce de Leon, J. 2004 *Gen. Rel. Grav.* **36** 1335. Ponce de Leon, J. 2007 gr-qc/0703094.

[11] Mashhoon, B., Liu, H., Wesson, P.S. 1994 *Phys. Lett. B* **331** 305. Mashhoon, B., Wesson, P.S. 2004 *Class. Quant. Grav.* **21** 3611. Mashhoon, B., Wesson, P.S. 2007 *Gen. Rel. Grav.* **39** 1403.

[12] Weinberg, S. 1980 *Rev. Mod. Phys.* **52** 515. Padmanabhan, T. 2003 *Phys. Rep.* **380** 235.

[13] Seahra, S.S., Wesson, P.S. 2001 *Gen. Rel. Grav.* **33** 1731.

[14] Youm, D 2001 *Mod. Phys. Lett. A* **16** 2371.

[15] Wesson, P.S., Mashhoon, B., Liu, H., Sajko, W.N. 1999 *Phys. Lett. B* **456** 34.

[16] Youm, D. 2000 *Phys. Rev. D* **62** 084002.

[17] Wesson, P.S. 2007 *Space-Time-Matter*, 2nd ed. (World Scientific: Singapore).

[18] Misner, C.W., Thorne, K.S., Wheeler, J.A. 1973 *Gravitation* (Freeman: San Francisco) 827. Carroll, S. 2004 *Spacetime and Geometry* (Addison-Wesley: San Francisco) 225.

[19] Wesson, P.S. 2003 *Gen. Rel. Grav.* **35** 307.

[20] Overduin, J.M., Wesson, P.S., Mashhoon, B. 2007 *Astron. Astrophys.* **473** 727. Wesson, P.S. 2005 Astron. Astrophys. **441** 41.

[21] Allman, B.E., Cimmino, A., Klein, A.G., Opat, G.I., Kaiser, H., Werner, S.A. 1992 *Phys. Rev. Lett.* **68** 2409. Allman, B.E., Cimmino, A., Klein, A.G., Opat, G.I., Kaiser, H., Werner, S.A. 1993 *Phys. Rev. A* **48** 1799. Rauch, H., Werner, S.A. 2000 *Neutron Interferometry* (Clarendon: Oxford).